\documentclass[prd,showpacs,showkeys,nofootinbib,floatfix,eqsecnum,
               fleqn,preprint,12pt,tightenlines]{revtex4-1} 

\usepackage{amsmath,amssymb,revsymb,graphicx,dcolumn}

\newcommand{\beq}{\begin{equation}}
\newcommand{\eeq}{\end{equation}}
\newcommand{\beqa}{\begin{eqnarray}}
\newcommand{\eeqa}{\end{eqnarray}}
\newcommand{\bsubeqs}{\begin{subequations}}
\newcommand{\esubeqs}{\end{subequations}}

\begin{document}

\begin{widetext}
\noindent Phys. Rev. D {\bf 101}, 064029 (2020) \hfill arXiv:1907.06547
%
%
\newline\vspace*{3mm}
\end{widetext}

\title{More on the regularized big bang singularity}

\vspace*{1mm}

\author{F.R. Klinkhamer}
\email{frans.klinkhamer@kit.edu}
\affiliation{Institute for Theoretical Physics,
Karlsruhe Institute of Technology (KIT),\\
76128 Karlsruhe, Germany\\}

\begin{abstract}
\vspace*{1mm}\noindent
The big bang singularity of the expanding-universe Friedmann solution
of the Einstein gravitational field equation can be regularized
by the introduction of a degenerate metric and a nonzero length scale $b$.
The result is a nonsingular bounce of the cosmic scale factor
with a contracting prebounce phase and an
expanding postbounce phase. The corresponding maximum values of
the curvature and the energy density occur at the moment of the
bounce and are proportional to powers of $1/b$. This article presents
a detailed calculation of the dynamics of such a nonsingular bounce.
In addition, a comparison is made between this nonsingular bounce
and the bounces of loop quantum cosmology and string cosmology.\\
\vspace*{-0mm}
\end{abstract}


\pacs{04.20.Cv, 98.80.Bp, 98.80.Jk}
\keywords{general relativity, big bang theory,
          mathematical and relativistic aspects of cosmology}

\maketitle

\section{Introduction}
\label{sec:Intro}

The Friedmann solution~\cite{Friedmann1922-1924,Weinberg1972,HawkingEllis1973}
of Einstein's gravitational field equation describes
an expanding universe, assumed to be homogeneous and isotropic.
The big bang singularity of the Friedmann solution
can be regularized~\cite{Klinkhamer2019-PRD} by the introduction
of a degenerate metric with a vanishing determinant on a 
three-dimensional submanifold of spacetime
and a nonzero length scale $b$.
The original big bang singularity
[at cosmic time coordinate $t=t_\text{bb}$ with vanishing
cosmic scale factor $a(t_\text{bb})=0$]
is replaced by a three-dimensional ``defect'' of spacetime
[the defect occurs at cosmic time coordinate $T=T_{B}$
and has a cosmic scale factor $a(T_{B})\ne0$,
for a new coordinate $T$ that is defined later].

The interpretation of the degenerate metric in
Ref.~\cite{Klinkhamer2019-PRD} as corresponding
to a spacetime defect~\cite{Klinkhamer2019-JPCS,KlinkhamerSorba2014}
is recalled below.  
At this moment, we only mention that we consider general relativity,
albeit in an extended version that allows for degenerate metrics
(see the last two paragraphs of Sec.~I in
Ref.~\cite{Klinkhamer2019-PRD} for further details).

Three follow-up papers~\cite{KlinkhamerWang2019-PRD,KlinkhamerWang2019-LHEP,%
KlinkhamerWang2019-metric-pert}
of the regularized-big-bang paper~\cite{Klinkhamer2019-PRD}
discuss certain phenomenological aspects of the resulting
nonsingular bouncing cosmology. See Ref.~\cite{IjjasSteinhardt2018}
for a review of the basic ideas of nonsingular
bouncing cosmology and an extensive list of references.

The calculations of the follow-up
papers~\cite{KlinkhamerWang2019-PRD,KlinkhamerWang2019-LHEP},
in particular, use an  auxiliary cosmic time coordinate  $\tau=\tau(T)$,
for which the  reduced field equations are nonsingular and, therefore,
directly accessible to numerical analysis.
These reduced field equations are, in fact,
ordinary differential equations (ODEs) and may be called
the $\tau$-ODEs.
But the auxiliary coordinate $\tau$ differs essentially
from the cosmic time coordinate $T$ that 
enters the metric. The corresponding reduced field equations
in terms of $T$ are singular ODEs and precisely the
singularities in these $T$-ODEs force the solution $a(T)$ to be
nonsingular, with a nonzero cosmic scale factor
$a(T_{B}) \ne 0$ at the moment of the cosmic bounce, $T=T_{B}$.

The goal of the present article is to carefully
study these singular $T$-ODEs,
in order to understand the dynamics of the time-symmetric nonsingular bounce. 
The outline is as follows.
In Sec.~\ref{sec:First-metric-Ansatz}, we recall the
\textit{Ansatz} for the metric from Ref.~\cite{Klinkhamer2019-PRD},
and discuss an advantage and a disadvantage of this \textit{Ansatz}.
In Sec.~\ref{sec:Second-metric-Ansatz}, we present a new \textit{Ansatz}
for the metric, which may or may not have a bounce,
depending on the dynamics.
In Sec.~\ref{sec:Bounce-solution-with-bcs-at-the-bounce}, we obtain analytic
and numerical results for this new metric with appropriate
boundary conditions at the bounce, where the main focus is
on establishing the smooth behavior of physical quantities at the bounce.
In Sec.~\ref{sec:Discussion}, we expand on the subtle issue
of boundary conditions (a calculation with initial conditions is presented in 
Appendix~\ref{app:Solutions-from-initial-conditions}).  
In Appendix~\ref{app:Comparison-with-loop-quantum-cosmology-and-string-cosmology},  
we compare our degenerate-metric bounce
with the bounces of loop quantum cosmology and string cosmology.

\section{First metric Ansatz}
\label{sec:First-metric-Ansatz}

For a modified spatially flat
Friedmann--Lema\^{i}tre--Robertson--Walker (FLRW) universe with
cosmic time coordinate $T$ and comoving
spatial Cartesian coordinates $\big\{X^{1},\,  X^{2},\, X^{3}\big\}$,
a relatively simple \textit{Ansatz} of a degenerate metric
is given by~\cite{Klinkhamer2019-PRD}%
\bsubeqs\label{eq:mod-FLRW}
\beqa\label{eq:mod-FLRW-ds2}
\hspace*{-2mm}
ds^{2}\,\Big|^\text{(type-1)}
&\equiv&
g_{\mu\nu}(X)\, dX^\mu\,dX^\nu \,\Big|^\text{(type-1)}
=
- \frac{T^{2}}{T^{2}+b^{2}}\;dT^{2}
+ \widetilde{a}^{\,2}(T) \;\delta_{kl}\,dX^k\,dX^l\,,
\\[2mm]
\hspace*{-2mm}
b^2 &>& 0\,,
\\[2mm]
\hspace*{-2mm}
\widetilde{a}(T)  &\in& \mathbb{R}\,,
\\[2mm]
\hspace*{-2mm}
T &\in& (-\infty,\,\infty)\,,\quad
X^k \in (-\infty,\,\infty)\,,
\eeqa
\esubeqs
where $b>0$ corresponds to the length scale of a spacetime defect
(cf. Refs.~\cite{Klinkhamer2019-JPCS,KlinkhamerSorba2014}
and references therein).
For definiteness, we call this previous metric \eqref{eq:mod-FLRW-ds2}
the ``type-1'' metric.
The metric from \eqref{eq:mod-FLRW-ds2} is degenerate, having
$\det\,g_{\mu\nu} = 0$ at $T=0$.  
We remark that the spacetime resulting from the
metric \eqref{eq:mod-FLRW}
is no longer independent of the choice of
foliation~\cite{Teitelboim1973,HojmanKucharTeitelboim1976},
as the hypersurface $T=0$ is singled out.
Incidentally, this degeneracy hypersurface can be shifted
to $T=T_{B}$ by replacing the
metric component $g_{00}$ in \eqref{eq:mod-FLRW-ds2}
by $-(T-T_{B})^{2}\big/\big((T-T_{B})^{2}+b^{2}\big)$.

We assume that the  matter content
is described by a homogeneous perfect fluid with
energy density $\rho_{M}(T)$ and pressure $P_{M}(T)$.
From the Einstein gravitational field equation~\cite{Weinberg1972}
and the metric \eqref{eq:mod-FLRW-ds2}, we then obtain
the dynamic equations for the variables $\widetilde{a}(T)$, $\rho_{M}(T)$,
and $P_{M}(T)$.
These equations are the energy-conservation equation of the matter,
the equation of state relating $P_{M}(T)$ to $\rho_{M}(T)$,
the modified first-order spatially flat Friedmann equation, and
the modified second-order spatially flat Friedmann equation:
\bsubeqs\label{eq:previous-bounce-mod-Feqs}
\beqa
\label{eq:previous-bounce-mod-Feq-rhoMprimeeq}
\hspace*{-0mm}&&
\rho^{\prime}_{M}+ 3\;\frac{\widetilde{a}^{\,\prime}}{\widetilde{a}}\;
\Big[\rho_{M}+P_{M}\Big] =0\,,
\\[2mm]
\label{eq:previous-bounce-mod-Feq-EOS}
\hspace*{-0mm}&&
P_{M} = P_{M} \big(\rho_{M}\big)\,,
\\[2mm]
\label{eq:previous-bounce-mod-Feq-1stFeq}
\hspace*{-0mm}&&
\left[1+ \frac{b^{2}}{T^{2}}\,\right]\,
\left( \frac{\widetilde{a}^{\,\prime}}{\widetilde{a}}\right)^{2}
= \frac{8\pi G_{N}}{3}\,\rho_{M}\,,
\eeqa
\beqa
\label{eq:previous-bounce-mod-Feq-2ndFeq}
\hspace*{-0mm}&&
\left[1+\frac{b^{2}}{T^{2}}\,\right]\,
\left(\frac{\widetilde{a}^{\,\prime\prime}}{\widetilde{a}}+
\frac{1}{2}\,\left( \frac{\widetilde{a}^{\,\prime}}{\widetilde{a}}\right)^{2}
\right)
-\frac{b^{2}}{T^{3}}\,\frac{\widetilde{a}^{\,\prime}}{\widetilde{a}}
=
-4\pi G_{N}\,P_{M}\,,
\eeqa
\esubeqs
where the prime stands for differentiation with respect to $T$.

The ODEs \eqref{eq:previous-bounce-mod-Feq-1stFeq}
and \eqref{eq:previous-bounce-mod-Feq-2ndFeq}
reproduce, in the formal limit $b\to 0$,
the standard Friedmann equations~\cite{Weinberg1972}.
The modified Friedmann equations \eqref{eq:previous-bounce-mod-Feq-1stFeq}
and \eqref{eq:previous-bounce-mod-Feq-2ndFeq} can, in fact,
be rewritten as the standard Friedmann equations with an
additional effective energy density $\rho_\text{defect}$
and an additional effective pressure $P_\text{defect}$,
both proportional to $-b^2/(b^2+T^2)$.
For a bounce solution~\cite{Klinkhamer2019-PRD}
with finite values of $\rho_{M}$ and $P_{M}$
at the moment of the bounce ($T=0$), the total effective energy density
$\rho_\text{total} \equiv \rho_{M}+\rho_\text{defect}$ and 
the total effective pressure $P_\text{total} \equiv P_{M}+P_\text{defect}$
violate the Null Energy Condition over a finite time interval $I_{B}$ 
around $T=0$: $\big[\rho_\text{total} + P_\text{total}\big]_{T \in I_{B}} < 0$.
We remark also that,
if $\widetilde{a}^{\,\prime}(T)/\widetilde{a}(T)$
were to vanish at a cosmic time $T=T_\text{stop}\ne 0$,
this would require a vanishing matter energy density,  
$\rho_{M}(T_\text{stop})=0$, according to \eqref{eq:previous-bounce-mod-Feq-1stFeq}.

The advantage of the metric \eqref{eq:mod-FLRW-ds2}
is that it takes the standard FLRW form,
\beqa\label{eq:standard-FLRW-ds2}
ds^{2}\,\Big|^\text{(type-1,\;$\tau$-coord.)}
&=&
- d\tau^{2}+ \widehat{a}^{\,2}(\tau)\;\delta_{kl}\,dX^k\,dX^l\,,
\eeqa
if we replace the coordinate $T$ by the coordinate $\tau$,
which is defined as follows:%
\beqa
\label{eq:mod-FLRW-tau-def}
\hspace*{-0mm}
\tau(T)&=&
\begin{cases}
 + \sqrt{b^{2}+T^{2}}\,,    & \;\;\text{for}\;\; T \geq 0\,,
 \\[2mm]
 - \sqrt{b^{2}+T^{2}}\,,    & \;\;\text{for}\;\; T \leq 0\,,
\end{cases}
\eeqa
where $\tau=-b$ and $\tau=b$
correspond to the single point $T=0$ on the cosmic time axis.
The coordinate transformation \eqref{eq:mod-FLRW-tau-def}
is noninvertible (two different $\tau$ values for the single
value $T=0$) and is not a diffeomorphism.
This implies that the differential structure of the
spacetime manifold with metric \eqref{eq:mod-FLRW-ds2}
differs from the differential structure of the
spacetime manifold  with metric \eqref{eq:standard-FLRW-ds2};
see Ref.~\cite{KlinkhamerSorba2014} for an extensive discussion.
For practical calculations~\cite{KlinkhamerWang2019-PRD,%
KlinkhamerWang2019-LHEP,KlinkhamerWang2019-metric-pert},
the metric \eqref{eq:standard-FLRW-ds2}
is to be preferred, as that metric is relatively simple and
the corresponding $\tau$-ODEs nonsingular.

But, with the different differential structure from
\eqref{eq:mod-FLRW-ds2} and \eqref{eq:standard-FLRW-ds2},
the actual study of the bounce at $T=0$ requires  
the $T$-ODEs \eqref{eq:previous-bounce-mod-Feqs}.
The disadvantage of the metric \textit{Ansatz} \eqref{eq:mod-FLRW-ds2}, then,
is that it explicitly depends on the coordinate $T$, as do the
corresponding ODEs \eqref{eq:previous-bounce-mod-Feq-1stFeq}
and \eqref{eq:previous-bounce-mod-Feq-2ndFeq}.
It would be preferable to have a metric
that depends only on the scale factor and its derivatives.
Another desirable property of this new metric
would be that the appearance of a bounce is not
hardwired into the metric \textit{Ansatz}
but that the bounce occurs dynamically.

\section{Second metric Ansatz}
\label{sec:Second-metric-Ansatz}

We now present another metric \textit{Ansatz}
for a modified spatially flat FLRW universe,
with the metric depending only on the scale factor and its derivatives,
apart from two constants ($b$ and $a_{B}$).
For definiteness, we call this new metric the ``type-2'' metric.
Specifically, the new metric reads
\bsubeqs\label{eq:NEW02-mod-FLRW}
\beqa\label{eq:NEW02-mod-FLRW-ds2}
\hspace*{-0mm}
ds^{2}\,\Big|^\text{(type-2)}
&=&
- \frac{\big[a(T)-a_{B}\big]^{2}}
       {\big[a(T)-a_{B}\big]^{2}+b^{2}\,\big[a'(T)/2\big]^{2}}\;dT^{2}
+ a^{\,2}(T)
\;\delta_{kl}\,dX^k\,dX^l\,,
\\[2mm]\hspace*{-0mm}
b^2 &>& 0\,,
\\[2mm]\hspace*{-0mm}
a_{B} &>& 0\,,
\\[2mm]\hspace*{-0mm}
a(T)  &\in& \mathbb{R}\,,
\\[2mm]\hspace*{-0mm}
\hspace*{-0mm}
T &\in& (-\infty,\,\infty)\,,
\quad
X^k \in (-\infty,\,\infty)\,,
\eeqa
\esubeqs
where the prime stands, again, for differentiation with respect to $T$.
With the metric \eqref{eq:NEW02-mod-FLRW},
a bounce occurs at $T=T_{B}$ if%
\bsubeqs\label{eq:NEW02-mod-FLRW-bounce-conditions}
\beqa
a(T_{B}) &=&   a_{B}\,,   
\\[2mm]
a'(T_{B}) &=& 0\,.
\eeqa
\esubeqs
Whether or not
the conditions \eqref{eq:NEW02-mod-FLRW-bounce-conditions} are fulfilled
depends on the dynamics and the boundary conditions (see below).
An explicit realization of the bounce behavior is given by
\beq
\label{eq:a-bounce-behavior}
a(T) \sim a_{B} +c_{2}\,(T-T_{B})^{2} \,,
\eeq
for a nonzero constant $c_{2}$.

We observe that close to a bounce,
with $a(T) \sim a_{B} +c_{2}\,T^{2}$  for $T_{B}=0$ and  $c_{2} \ne 0$,
the $g_{00}$ component from \eqref{eq:NEW02-mod-FLRW-ds2} reduces to the
expression $-T^{4}/(T^{4}+b^{2}\,T^{2})$,
which has essentially the
same structure as the $g_{00}$ component of the
type-1 metric \eqref{eq:mod-FLRW-ds2}.
In principle, it is also possible to consider a metric \textit{Ansatz}
without explicit mention of $a_{B}$ \big(an example would be
$g_{00}= -(a')^{2}\big/\big[(a')^{2}+b^{2}\,(a'')^{2}\big]$\big),
but the \textit{Ansatz}
\eqref{eq:NEW02-mod-FLRW-ds2} suffices
for the purpose of studying the bounce dynamics.
Note that the type-2 metric \textit{Ansatz}
\eqref{eq:NEW02-mod-FLRW} differs from the
type-1 metric \textit{Ansatz} \eqref{eq:mod-FLRW} 
because, with the respective dynamic equations,
the type-2 metric may or may not
have a bounce, whereas the type-1 metric always has a bounce at $T=0$,
as long as $\rho_{M} > 0$
(see Appendix~\ref{subapp:Solution-without-bounce}
for a direct comparison of the two metrics).

Just as in Sec.~\ref{sec:First-metric-Ansatz}, we
assume that the matter content is given by
a homogeneous perfect fluid.
From the Einstein gravitational field equation~\cite{Weinberg1972}
and the new metric \eqref{eq:NEW02-mod-FLRW-ds2}, we then obtain
the dynamic equations for the variables $a(T)$, $\rho_{M}(T)$,
and $P_{M}(T)$.
These equations are the energy-conservation equation of the matter,
the equation of state of the matter,
the (new) modified first-order spatially flat Friedmann equation,
and the (new) modified second-order spatially flat Friedmann equation:%
\bsubeqs\label{eq:NEW02-mod-Feqs}
\beqa
\label{eq:NEW02-mod-Feq-rhoMprimeeq}
\hspace*{-4.0mm}
&&
\rho^{\prime}_{M}+ 3\,\frac{a^{\prime}}{a}\,\Big[\rho_{M}+P_{M}\Big] =0\,,
\\[2mm]
\label{eq:NEW02-mod-Feq-EOS}
\hspace*{-4.0mm}
&&
P_{M} = P_{M} \big(\rho_{M}\big)\,,
\\[2mm]
\label{eq:NEW02-mod-Feq-1stFeq}
\hspace*{-4.0mm}
&&
\left[1+ \frac{b^{2}}{4}\,\frac{a^{2}}{\big[a-a_{B}\big]^{2}}\,
\left( \frac{a^{\prime}}{a}\right)^{2}\,\right]\,
\left( \frac{a^{\prime}}{a}\right)^{2}
= \frac{8\pi G_{N}}{3}\,\rho_{M}\,,
\\[2mm]
\label{eq:NEW02-mod-Feq-2ndFeq}
\hspace*{-4.0mm}
&&
\left[1+ \frac{1}{2}\,\frac{b^{2}\,\big[a'\big]^{2}}{\big[a-a_{B}\big]^{2}}
\right]\,
\frac{a^{\prime\prime}}{a}
+\frac{1}{2}\,\left( \frac{a^{\prime}}{a}\right)^{2}
+\frac{1}{8}\,b^{2}\,\frac{a^{2}\,\big[a+a_{B}\big]}{\big[a-a_{B}\big]^{3}}
\, \left( \frac{a^{\prime}}{a}\right)^{4}
= - 4\pi G_{N}\,P_{M}\,.
\eeqa
\esubeqs
We have the following remarks:
\begin{enumerate}
  \item
The ODEs \eqref{eq:NEW02-mod-Feq-1stFeq} and \eqref{eq:NEW02-mod-Feq-2ndFeq}
reproduce, in the formal limit $b\to 0$,
the standard Friedmann equations~\cite{Weinberg1972}.
\item
Precisely the $b^{2}$ terms of the
ODEs \eqref{eq:NEW02-mod-Feq-1stFeq} and \eqref{eq:NEW02-mod-Feq-2ndFeq}
contain various powers
of the factor $a^{\prime}(T)\big/\big[a(T)-a_{B}\big]$,
which is singular at $T=T_{B}$
for $a(T)$ from \eqref{eq:a-bounce-behavior}.
\item
The singular $b^{2}$ term in \eqref{eq:NEW02-mod-Feq-1stFeq}
allows for a nontrivial bounce solution at $T=T_{B}$, with
$a(T)$ from \eqref{eq:a-bounce-behavior}
and $\rho_{M}(T) \sim r_{0} +r_{2}\,(T-T_{B})^{2}$
for positive $a_{B}$ and $r_{0}$;
see Sec.~\ref{subsec:Analytic-results} for further details.
\item
If $a^{\,\prime}(T)$ were to vanish
at a cosmic time $T=T_\text{stop}\ne T_{B}$
with $a(T_\text{stop})\ne 0$ and $a(T_\text{stop})\ne a_{B}$,
this would require a vanishing energy density, $\rho_{M}(T_\text{stop})=0$,
according to \eqref{eq:NEW02-mod-Feq-1stFeq}.
\item
It can be verified that
the second-order ODE \eqref{eq:NEW02-mod-Feq-2ndFeq} follows from
the first-order ODEs \eqref{eq:NEW02-mod-Feq-rhoMprimeeq}
and \eqref{eq:NEW02-mod-Feq-1stFeq}; 
see the discussion in Sec.~15.1 of Ref.~\cite{Weinberg1972}   
for the case of the standard Friedmann equations.
\item
The ODEs \eqref{eq:NEW02-mod-Feq-rhoMprimeeq},
\eqref{eq:NEW02-mod-Feq-1stFeq}, and \eqref{eq:NEW02-mod-Feq-2ndFeq}
are invariant under the rescaling $a(T)\to \zeta\,a(T)$ and
$a_{B}\to \zeta\,a_{B}$ for $\zeta \in \mathbb{R}\backslash\{0\}$,
and also under time reversal $(T-T_{B}) \to -(T-T_{B})$
if $a$, $\rho_{M}$, and $P_{M}$ are even functions of $T-T_{B}$.
\end{enumerate}

From now on, we assume that the matter content is
described by a homogeneous perfect fluid with
a \emph{constant} equation-of-state parameter,
\beq
\label{eq:constant-wM}
W_{M}(T) \equiv P_{M}(T)/\rho_{M}(T) = w_{M} =\text{constant}\,.
\eeq
Furthermore, we use reduced-Planckian units, with
\beq
\label{eq:reduced-Planckian-units}
8\pi G_{N}=c=\hbar=1\,,
\eeq
and take the following model parameters:
\bsubeqs\label{eq:model-parameters}
\beqa\label{eq:model-parameter-b}
b &=& 1\,,
\\[2mm]
\label{eq:model-parameter-wM}
w_{M} &=& 1\,,
\eeqa
\esubeqs
where this particular choice for $w_{M}$ aims
at avoiding instabilities in the prebounce phase
(see, e.g., Sec.~IV of Ref.~\cite{IjjasSteinhardt2018}
and references therein).
In order to compare with previous results, we choose
the following value for the 
bounce scale factor in \eqref{eq:NEW02-mod-FLRW}
and \eqref{eq:NEW02-mod-Feqs}    
\bsubeqs\label{eq:bounce-parameters}
\beqa
\label{eq:bounce-parameter-aB}
a_{B}&=& 1\,,
\eeqa
and take the following value for the bounce time 
\beqa
\label{eq:bounce-parameter-TB}
T_{B}&=& 0\,,
\eeqa
\esubeqs
but the value of $a_{B}$ can be arbitrarily rescaled
and the value of $T_{B}$ can be arbitrarily shifted.

\section{Bounce solution with boundary conditions at the bounce}
\label{sec:Bounce-solution-with-bcs-at-the-bounce}

In this section, we discuss the solution of the
ODEs \eqref{eq:NEW02-mod-Feqs}
with appropriate boundary conditions at the bounce.
Specifically, we take the following boundary conditions:
\bsubeqs\label{eq:boundary-conditions-at-bounce}
\beqa
a(0) &=&  1\,,
\\[2mm]
a'(0) &=& 0\,,
\\[2mm]
\rho_{M}(0) &>& 0\,,
\eeqa
\esubeqs
where we have used the values $T_{B}=0$ and $a_{B}=1$
from \eqref{eq:bounce-parameters}.

In Sec.~\ref{sec:Discussion}, we give 
a general discussion of the issue of boundary conditions and,
in Appendix~\ref{app:Solutions-from-initial-conditions},
we obtain solutions with initial conditions,
which may or may not result in a bounce solution.

\subsection{Analytic results}
\label{subsec:Analytic-results}

We start from the analytic solution~\cite{Weinberg1972}
of \eqref{eq:NEW02-mod-Feq-rhoMprimeeq} for the case of a
constant equation-of-state parameter $w_{M}$
as defined by \eqref{eq:constant-wM}  
\bsubeqs\label{eq:rhoMsol-r0-positive}
\beqa
\label{eq:rhoMsol}
\rho_{M}(a) &=& r_{0}\;a^{-3\,\left[1+w_{M}\right]}\,,
\\[2mm]
\label{eq:r0-positive}
r_{0} &>& 0\,,
\eeqa
\esubeqs
where $a(T)$ is assumed to be positive and to have
boundary conditions \eqref{eq:boundary-conditions-at-bounce}.
The resulting ODEs \eqref{eq:NEW02-mod-Feq-1stFeq}
and \eqref{eq:NEW02-mod-Feq-2ndFeq} near $T=T_{B}=0$
are approximately solved by a truncated power series,
\beq
\label{eq:a-power-series}
a(T) = 1+ \sum_{n=1}^{N} a_{2\,n}\,\big(T/b\big)^{\,2\,n}\,,
\eeq
for appropriate values of the coefficients $a_{2\,n}$.
With $N=4$, we obtain the following coefficients
(taking the positive root for $a_{2}$):
\bsubeqs\label{eq:a2-a4-a6-a8}
\beqa
\hspace*{-10mm}
\label{eq:a2}
a_{2} &=&
\frac{1}{2\,{\sqrt{3}}}\;
b\,{\sqrt{r_{0}}}
\,,
\\[2mm]
\hspace*{-10mm}
a_{4} &=&
-\frac{1}{72\,{\sqrt{3}}}\;
\left(
6\,b\,{\sqrt{r_{0}}} +
    {\sqrt{3}}\,b^{2}\,r_{0}\,\left[ 1 + 3\,w_{M} \right]\right)
    \,,
\\[2mm]
\hspace*{-10mm}
a_{6} &=&
\frac{1}{25920}\;
b\,{\sqrt{r_{0}}}\,
   \Big( 156\,{\sqrt{3}} + 48\,b\,{\sqrt{r_{0}}}\,
       \left[ 1 + 3\,w_{M} \right]
       +       {\sqrt{3}}\,b^{2}\,r_{0}\,
       \left[ 31 + 132\,w_{M} + 117\,{w_{M}}^{2} \right]  \Big)
      \,,
\nonumber\\[1mm] \hspace*{-10mm}&&
\eeqa
\beqa
 \hspace*{-10mm}
a_{8} &=&
-\frac{1}{362880\, {\sqrt{3}}}\;
\;b\,{\sqrt{r_{0}}}\;
      \Big( 1044 - 72\,{\sqrt{3}}\,b\,{\sqrt{r_{0}}}\,
         \left[ 1 + 3\,w_{M} \right]
         +   9\,b^{2}\,r_{0}\,\left[ 29 + 120\,w_{M} + 99\,{w_{M}}^{2} \right]
\nonumber\\[1mm]
\hspace*{-10mm}&&
           +         2\,{\sqrt{3}}\,b^3\,{r_{0}}^{3/2}\,
         \left[ 77 + 405\,w_{M} + 621\,{w_{M}}^{2} +
           297\,{w_{M}}^3 \right]  \Big)   \,,
\eeqa
\esubeqs
where $r_{0}$ in the expressions \eqref{eq:a2-a4-a6-a8}
corresponds to $8\pi G_{N}\,\rho_{0}$ with mass dimension $2$,
so that $b\,\sqrt{r_{0}}$ is dimensionless.  
For $b=1$, $w_{M} = 1$, and $r_{0}=1/3$, we have
the following numerical values:
\beq
\label{eq:a2-a4-a6-a8-num}
\big\{a_{2},\, a_{4},\,  a_{6},\,  a_{8}\big\}
\,\Big|_{b=1,\; w_{M} = 1,\; r_{0}=1/3}
\approx
\big\{0.166667,\, -0.0462963,\,  0.0120885,\,  -0.0022352\big\}\,,
\eeq
which suggests an alternating series with a finite
radius of convergence.

Returning to the $a_{2}$ solution \eqref{eq:a2},
the relevant equation is the series expansion
of the modified first-order spatially flat Friedmann equation
\eqref{eq:NEW02-mod-Feq-1stFeq}, which reads
\beq
\label{eq:NEW02-mod-Feq-1stFeq-series}
0= 4\,(a_{2})^{2}/b^{2} - r_{0}/3 + \text{O}(T^{2})\,,
\eeq
for $8\pi G_{N}=1$ and $r_{0} > 0$.
As said, we have chosen the positive root for $a_{2}$ and postpone
further discussion to Sec.~\ref{sec:Discussion}.

The Ricci curvature scalar
$R(x) \equiv g^{\nu\sigma}(x)\,g^{\mu\rho}(x)\,R_{\mu\nu\rho\sigma}(x)$
and the Kretschmann curvature scalar
$K(x) \equiv R^{\mu\nu\rho\sigma}(x)\,R_{\mu\nu\rho\sigma}(x)$
are readily evaluated for the metric \eqref{eq:NEW02-mod-FLRW-ds2}.
The resulting expressions are functionals of the \textit{Ansatz}
function $a(T)$ and its derivatives. Inserting the truncated power
series \eqref{eq:a-power-series}, we obtain
\beqa
\label{eq:R-power-series}
R(T) = \frac{1}{b^{2}}\,\sum_{n=0}^{N'} R_{2\,n}\,\big(T/b\big)^{\,2\,n}\,,
\\[2mm]
\label{eq:K-power-series}
K(T) = \frac{1}{b^{4}}\,\sum_{n=0}^{N''} K_{2\,n}\,\big(T/b\big)^{\,2\,n}\,.
\eeqa
With the $a$-coefficients from \eqref{eq:a2-a4-a6-a8}, we get
\bsubeqs\label{eq:R0-R2-R4}
\beqa
\label{eq:R0}
\hspace*{-5mm}
R_{0} &=&
b^{2}\,r_{0}\, \left( 1 - 3\, w_{M} \right)
\,,
\\[2mm]
\hspace*{-5mm}
R_{2} &=&
-\frac{1 }{2}\;
{\sqrt{3}}\; b^3\, {r_{0}}^{3/2}\,
\left( 1 - 2\, w_{M} - 3\, {w_{M}}^{2} \right)
\,,
\\[2mm]
\hspace*{-5mm}
R_{4} &=&
\frac{1 }{24}\;
b^3\, {r_{0}}^{3/2}\, \left( 1 - 2\, w_{M} - 3\, {w_{M}}^{2} \right)
\, \left( 2\, {\sqrt{3}} + b\, {\sqrt{r_{0}}}\,
              \left[13 + 12\, w_{M} \right]  \right)  \,,
\eeqa
\esubeqs
and
\bsubeqs\label{eq:K0-K2-K4}
\beqa
\hspace*{-5mm}
K_{0} &=&
\frac{1}{3}\;
b^{4}\, {r_{0}}^{2}\, \left( 5 + 6\, w_{M} +
    9\, {w_{M}}^{2} \right)
\,,
\\[2mm]
\hspace*{-5mm}
K_{2} &=&
-\frac{1 }{{\sqrt{3}}} \;
b^5\, {r_{0}}^{5/2}\, \left( 5 + 11\, w_{M}
                  + 15\, {w_{M}}^{2}
                  + 9\, {w_{M}}^3 \right)\,,
\\[2mm]
\hspace*{-5mm}
K_{4} &=&
\frac{1}{36}\;
b^5\, {r_{0}}^{5/2}\,
\left( 5 + 11\, w_{M} + 15\, \
{w_{M}}^{2} + 9\, {w_{M}}^3 \right)
\, \left( 2\, {\sqrt{3}} + b\, {\sqrt{r_{0}}}\,
\left[ 22 + 21\, w_{M} \right]  \right)\,.
\eeqa
\esubeqs
%
%
\begin{figure}[t!]
\vspace*{-0mm}
\begin{center}
\includegraphics[width=1\textwidth]{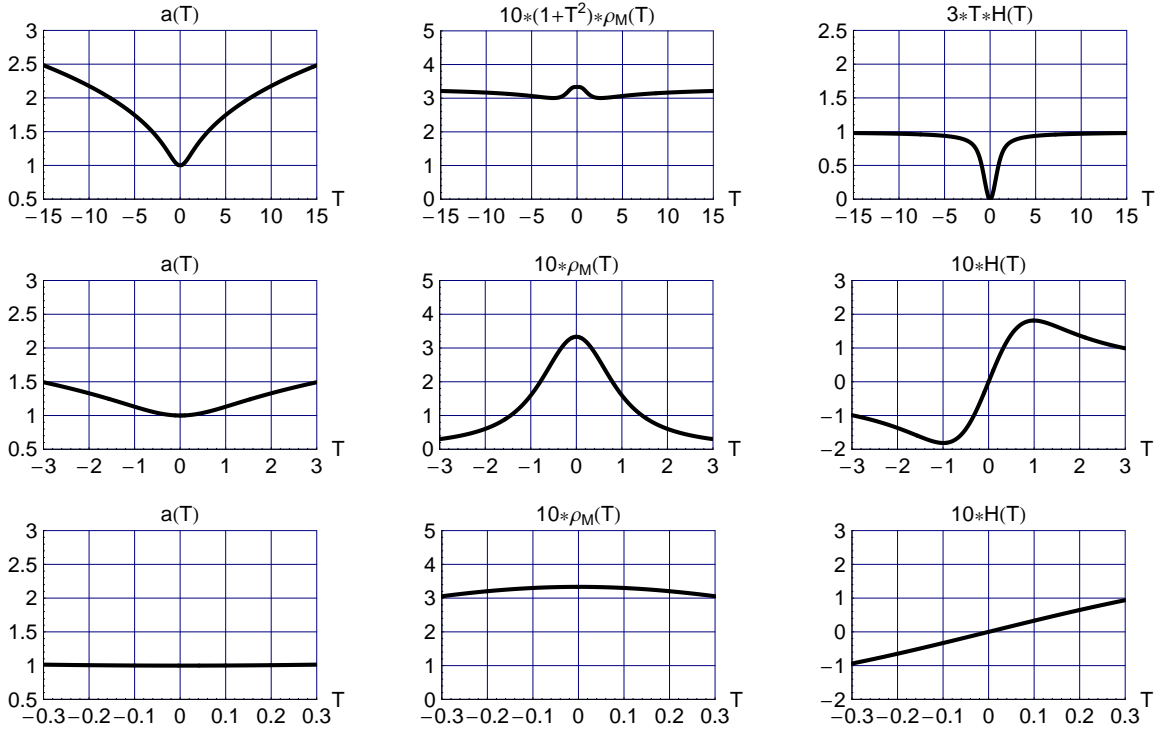}
\end{center}
\vspace*{-5mm}
\caption{Solution of the ODEs
\eqref{eq:NEW02-mod-Feq-rhoMprimeeq} and \eqref{eq:NEW02-mod-Feq-1stFeq}
for a constant equation-of-state parameter $w_{M}$ from \eqref{eq:constant-wM}
and reduced-Planckian units \eqref{eq:reduced-Planckian-units}.
The model parameters are $b=1$ and $w_{M} = 1$.
The boundary conditions at $T=T_{B}=0$ are
$a(0)= a_{B} = 1$ and $\rho_{M}(0)=r_{0}$ for $r_{0}=1/3$.
The approximate analytic solution is shown
over $T \in (-\Delta T,\,\Delta T)$ and
the numerical solution over
$T \in [-T_\text{max},\,-\Delta T] \cup [\Delta T,\,T_\text{max}]$,
with $\Delta T=1/10$ and $T_\text{max}=15$.
Specifically, the approximate analytic solution for $a(T)$ is given
by \eqref{eq:a-power-series} and \eqref{eq:a2-a4-a6-a8} for $N=4$,
while the analytic solution for $\rho_{M}(T)$
follows from \eqref{eq:rhoMsol-r0-positive}.
The numerical solution is obtained from
the ODEs \eqref{eq:NEW02-mod-Feq-rhoMprimeeq}
and \eqref{eq:NEW02-mod-Feq-1stFeq} with boundary conditions
at $T=\pm \Delta T$ from the approximate analytic solution.
Shown, on the top row, are the dynamic functions $a(T)$ and
$\rho_{M}(T)$, together with the corresponding
Hubble parameter $H(T) \equiv [d a(T)/d T]/a(T)$.
The middle and bottom rows zoom in on the bounce at $T=0$.
In the middle panel of the top row, $10\,\rho_{M}(T)$ is scaled by a
further factor $\big(1+T^{2}\big)$, in order to display the asymptotic behavior
$\rho_{M}(T) \propto  1/T^{2}$ as $|T|\to\infty$.
Similarly, in the right panel of the top row, $H(T)$ is scaled by a
factor $3\,T$, in order to display the asymptotic
behavior $H(T) \sim (1/3)\,T^{-1}$.
}
\label{fig:fig1}
\vspace*{0mm}
\end{figure}

The perturbative results \eqref{eq:R0-R2-R4}  
suggest that $R(T)=0$ for the case of relativistic matter with
$w_{M}=1/3$, just as for the standard FLRW universe~\cite{Weinberg1972}.
The numerical values of the coefficients
\eqref{eq:R0-R2-R4} and \eqref{eq:K0-K2-K4}
for $b=1$, $w_{M} = 1$, and $r_{0}=1/3$ are
\bsubeqs\label{eq:R0-R2-R4-K0-K2-K4-num}
\beqa
\label{eq:R0-R2-R4-num}
\big\{R_{0},\, R_{2},\,  R_{4}\big\}
\,\Big|_{b=1,\; w_{M} = 1,\; r_{0}=1/3}
&\approx&
\big\{-0.666667,\,  0.666667,\,  -0.574074\big\}\,,
\eeqa
\beqa
\label{eq:K0-K2-K4-num}
\big\{K_{0},\, K_{2},\,  K_{4}\big\}
\,\Big|_{b=1,\; w_{M} = 1,\; r_{0}=1/3}
&\approx&
\big\{0.740741,\,  -1.48148,\,  2.01646\big\}\,.
\eeqa
\esubeqs

For completeness, we also give the asymptotic solution
for $|T|\to\infty$,
\bsubeqs\label{eq:a-rhoM-asymptotic solution}
\beqa
a_\text{asymp}(T) &\sim& a_{\infty}\,(T^{2})^{p/2}\,,
\\[2mm]
\rho_{M\,\text{asymp}}(T) &\sim&
r_{\infty}\;\big[a_\text{asymp}(T)\big]^{-3\,\left[1+w_{M}\right]}\,,
\\[2mm]
a_{\infty} &=& \left(\frac{r_{\infty}}{3\,p^{2}}\right)^{p/2}\,,
\\[2mm]
p &=& \frac{2}{3}\;\frac{1}{1+w_{M}}\,,
\eeqa
\esubeqs
where the constant $r_{\infty}$ depends indirectly on the
constant $r_{0}$ from \eqref{eq:rhoMsol-r0-positive}.

\begin{figure}[t]
\vspace*{-0mm}
\begin{center}
\includegraphics[width=1\textwidth]{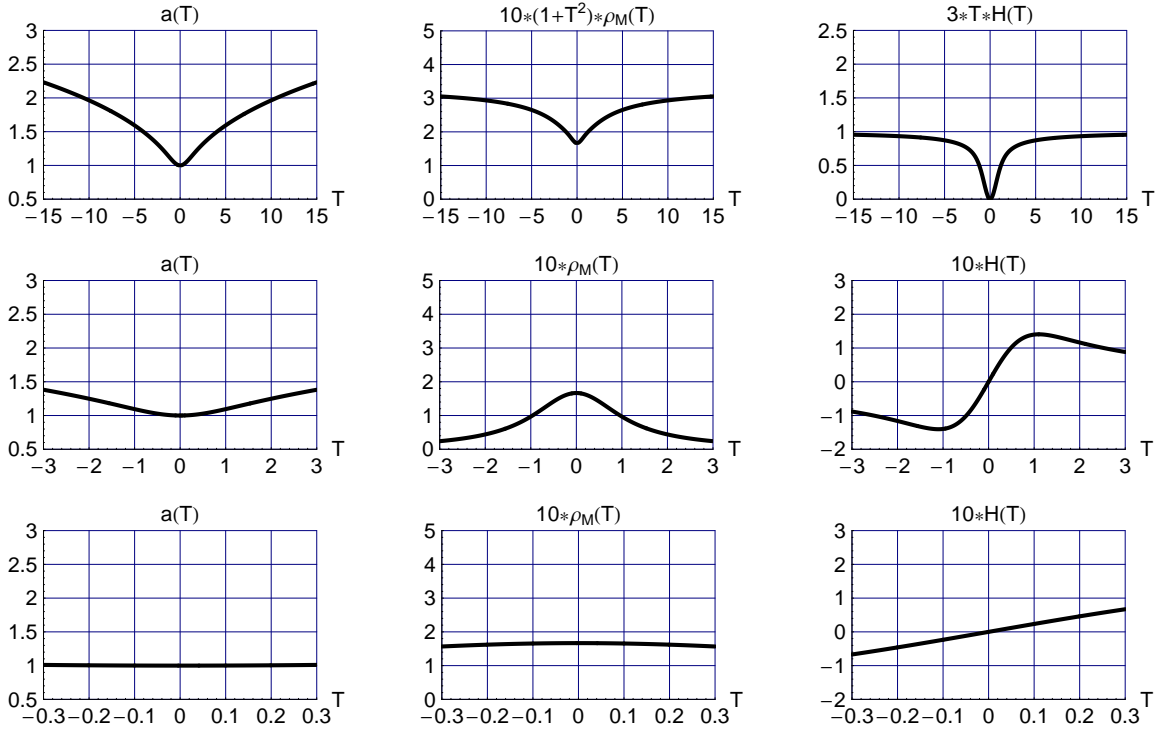}
\end{center}
\vspace*{-8mm}
\caption{Same as Fig.~\ref{fig:fig1},
but now with boundary condition $\rho_{M}(0)=r_{0}=1/6$.
}
\label{fig:fig2}
\vspace*{0mm}
\end{figure}

In closing, we remark that we have seen that
the solution $\overline{a}(T)$ of the ODEs \eqref{eq:NEW02-mod-Feqs}
can be expanded perturbatively around $T=T_{B}=0$.
The obtained Taylor series is
characterized by the numerical value of $r_{0}$,
as the values of $T_{B}$ and $a_{B}$ can be changed arbitrarily
(here, they are taken as $T_{B}=0$ and $a_{B}=1$).
The asymptotic solution for $a(T)$ has also been found to depend
indirectly on $r_{0}$.
Similar results hold for the solution~\cite{Klinkhamer2019-PRD}
from the type-1 metric of Sec.~\ref{sec:First-metric-Ansatz},
but, here, we have focused on the more general
type-2 metric of Sec.~\ref{sec:Second-metric-Ansatz}.

\subsection{Numerical results}
\label{subsec:Numerical-results}

We obtain a close approximation of the exact solution
$\overline{a}(T)$ of the ODEs \eqref{eq:NEW02-mod-Feqs},
for a constant equation-of-state parameter $w_{M}$
from \eqref{eq:constant-wM} and with
boundary conditions \eqref{eq:boundary-conditions-at-bounce},
by using the truncated power series \eqref{eq:a-power-series}
for a small enough interval around $T=T_{B}=0$
and by solving the ODEs
\eqref{eq:NEW02-mod-Feq-rhoMprimeeq} and \eqref{eq:NEW02-mod-Feq-1stFeq}
numerically for $T$ values outside this central interval.
The ODE \eqref{eq:NEW02-mod-Feq-1stFeq} is singular at $T=0$,
if $a(0)=a_{B}=1$,
but this value $T=0$ lies \emph{outside} the intervals used for the
numeric calculation. Specifically, we choose
the power-series interval $[-\Delta T,\,\Delta T]$
and truncate the series \eqref{eq:a-power-series} at $N=4$.   
In principle, we must
take $\Delta T \to 0$ and $N\to\infty$. We leave a detailed study
of the numerical convergence properties to the future, as well as
an analytic calculation (if at all possible)
of the
radius of convergence of the power series
corresponding to \eqref{eq:a-power-series}.
For the moment, we have just compared the results for different values of
$\Delta T$ (specifically, $1/10$ or $1/100$) and
different values of $N$ (specifically, $4$ or $8$),
and find the results to be reasonably stable.

\begin{figure}[t]  
\vspace*{-0mm}
\begin{center}   
\includegraphics[width=1\textwidth]{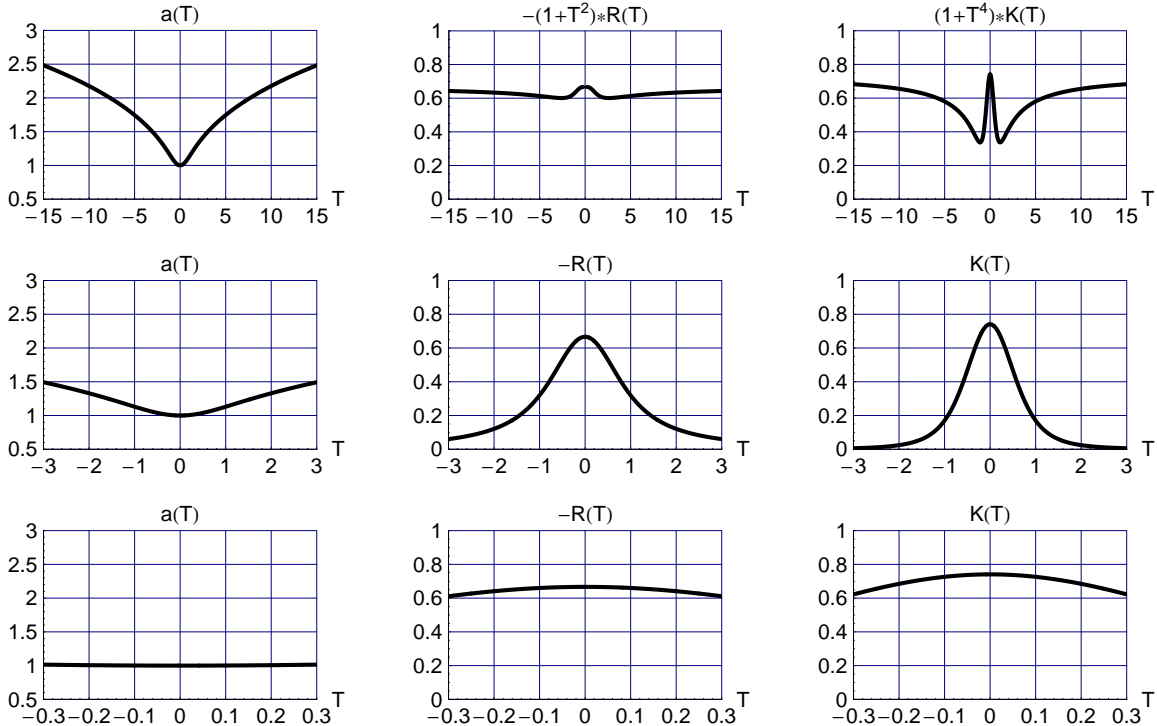}
\end{center}
\vspace*{-8mm}
\caption{Ricci curvature scalar $R(T)$ and Kretschmann curvature scalar $K(T)$
for the solution $a(T)$ of Fig.~\ref{fig:fig1}.
In the middle panel of the top row, $-R(T)$ is scaled by a factor
$\big(1+T^{2}\big)$, in order to display the asymptotic behavior
$-R(T) \propto  1/T^{2}$.
Similarly, in the right panel of the top row, $K(T)$ is scaled by a factor
$\big(1+T^{4}\big)$, in order to display the asymptotic behavior
$K(T) \propto  1/T^{4}$.}
\label{fig:fig3}
\end{figure}

For the numerical solution of the ODEs \eqref{eq:NEW02-mod-Feqs},
we can focus on the modified first-order Friedmann
ODE \eqref{eq:NEW02-mod-Feq-1stFeq}. The reason is that
\eqref{eq:NEW02-mod-Feq-rhoMprimeeq} already has the analytic solution
\eqref{eq:rhoMsol} and that, as mentioned in the fifth remark below
\eqref{eq:NEW02-mod-Feq-2ndFeq},
the second-order Friedmann ODE \eqref{eq:NEW02-mod-Feq-2ndFeq}
follows from the first-order ODEs \eqref{eq:NEW02-mod-Feq-rhoMprimeeq}
and \eqref{eq:NEW02-mod-Feq-1stFeq}.

\begin{figure}[t]  
\vspace*{-0mm}
\begin{center}   
\includegraphics[width=1\textwidth]{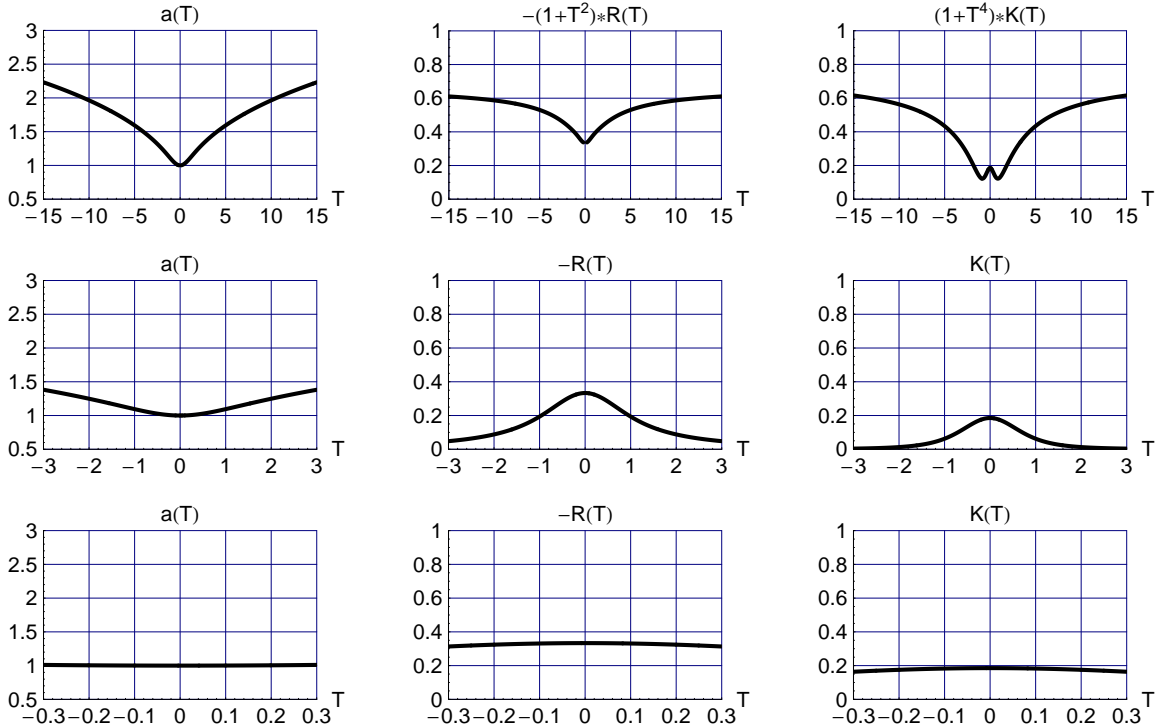}
\end{center}
\vspace*{-5mm}
\caption{Ricci curvature scalar $R(T)$ and Kretschmann curvature scalar $K(T)$
for the solution $a(T)$ of Fig.~\ref{fig:fig2}.
The scaling of $-R(T)$ and $K(T)$ in the top-row panels is the same as used in
Fig.~\ref{fig:fig3}.
}
\label{fig:fig4}
\vspace*{0mm}
\end{figure}

As to the numerical solution of the ODE \eqref{eq:NEW02-mod-Feq-1stFeq},
there are two subtleties. The first subtlety is that the
ODE \eqref{eq:NEW02-mod-Feq-1stFeq} is not directly
accessible to numerical analysis, as the equation is quadratic in
$\big[a'\big]^{2} \equiv S$. But we can obtain analytically
the positive root of this quadratic equation for $S$.
Then, we take the square root of $S$, with a minus sign of
the resulting expression for $a'$ in the prebounce phase and
a plus sign in the postbounce phase.
The second subtlety is that we do not numerically solve the
obtained ODEs which are linear in $a'$
(with different signs for the pre- and postbounce phases),
but rather numerically solve the
first derivative of these first-order ODEs. In this way, we obtain
a numerical solution with a reasonably accurate value of $a''(T)$ at
$T= \pm \Delta T$. In fact, we can check for the accuracy of the obtained
numerical solution $a_\text{num}(T)$ by evaluating
the residue of the  second-order ODE \eqref{eq:NEW02-mod-Feq-2ndFeq}.

Analytic and numerical results for $b=1$, $w_{M} = 1$, and $r_{0}=1/3$
are given in Fig.~\ref{fig:fig1}.
As noted in the last paragraph of Sec.~\ref{subsec:Analytic-results},
the solution is characterized by the numerical value of $r_{0}$,
for fixed values of the model parameters $b$ and $w_{M}$.
We give the results for a smaller numerical value of $r_{0}$
in Fig.~\ref{fig:fig2}, where the $\rho_{M}$ peak is found to be
lower and somewhat broader than the one in Fig.~\ref{fig:fig1}.
The corresponding plots for
the Ricci curvature scalar $R(T)$ and
the Kretschmann curvature scalar $K(T)$
are given in Figs.~\ref{fig:fig3} and \ref{fig:fig4}.

The top-right panels in Figs.~\ref{fig:fig1} and \ref{fig:fig2}
show the asymptotic behavior $H(T) \sim (1/3)\,T^{-1}$, which
results from the asymptotic behavior $a(T) \propto (T^{2})^{1/6}$,
as given by the analytic solution \eqref{eq:a-rhoM-asymptotic solution}.
The bottom rows in Figs.~\ref{fig:fig1} and \ref{fig:fig2}
illustrate the smooth behavior at the bounce $T=T_{B}=0$.
The smoothness of the bounce is
also evident from the behavior of
the Ricci curvature scalar $R(T)$
and the Kretschmann curvature scalar $K(T)$,
as obtained perturbatively in
\eqref{eq:R-power-series} and \eqref{eq:K-power-series}  
and shown on the bottom rows of Figs.~\ref{fig:fig3} and \ref{fig:fig4},

\section{Discussion}
\label{sec:Discussion}

We have presented, in Sec.~\ref{sec:Second-metric-Ansatz},
a new metric \textit{Ansatz}
for a modified spatially flat FLRW universe.
For the case of a constant equation-of-state parameter $w_{M}$
and with appropriate boundary conditions, we have obtained,
in Sec.~\ref{sec:Bounce-solution-with-bcs-at-the-bounce},
analytic results in a small interval around the cosmic time $T=T_{B}=0$
of the time-symmetric bounce, together with numerical results
at larger values of $|T|$ for the case of $w_{M}=1$.
The solution $\overline{a}(T)$ is regular at $T=0$
and appears to be well behaved for finite values of $|T|$.
The solution $\overline{a}(T)$
is characterized by the maximum energy density $\rho_{M}(T)$
of the matter, which occurs at $T=0$.
The dimensionless quantity for this maximum energy density is
denoted $r_{0}$, and the behavior of the analytic and numeric
solutions in Figs.~\ref{fig:fig1}
and \ref{fig:fig2} is controlled by $r_{0}$ only,
for fixed model parameters $b$ and $w_{M}$.

As mentioned in Sec.~\ref{subsec:Analytic-results},
the behavior of the $a(T)$ power series solution
\eqref{eq:a-power-series}
near $T=T_{B}=0$ has been chosen to be convex
($a_{2}>0$ with $a_{B}=1$).
In principle, it is also possible to have an
$a(T)$ solution near $T=T_{B}$
that is concave ($a_{2}<0$ with $a_{B}=1$),
so that there will be
cosmic times $T_{\pm}=T_{B} \pm \Delta T_\text{bb}$ with a
vanishing scale factor, $a(T_{\pm})=0$.
The different bounces, convex and concave at $T=T_{B}$,
result from different initial conditions at $T_\text{start} < T_{B}$.
Taking $a(T_\text{start}) > a_{B}$ and $a'(T_\text{start}) <0$
gives a convex bounce, with a contracting phase
for $T\in [T_\text{start},\,T_{B})$
and an expanding phase for $T\in (T_{B},\,\infty)$.
An explicit bounce solution from such
initial conditions is presented in
Appendix~\ref{subapp:Solution-with-bounce}.
Solutions with other initial conditions are obtained in
Appendix~\ref{subapp:Solution-without-bounce},
which illustrate the difference between the metrics of
Secs.~\ref{sec:First-metric-Ansatz} and \ref{sec:Second-metric-Ansatz}.

Up till now, we have not been specific as regards the numerical value
of the defect length scale $b$, apart from the fact that
this numerical value should not be too small,
so as to invalidate the approximate
applicability of Einstein's classical theory of gravity.
In Ref.~\cite{KlinkhamerWang2019-PRD}, we have obtained a
broad range of numerical $b$ values allowed by experiment.
A hint that the defect length scale $b$ may be close to the
Planck length is obtained in
Appendix~\ref{app:Comparison-with-loop-quantum-cosmology-and-string-cosmology}
by comparing  our degenerate-metric bounce
with the bounce of loop quantum cosmology~\cite{AshtekarSingh2006,%
AshtekarSingh2011,AshtekarBarrau2015}
(for completeness, we also compare with the bounce of string
cosmology~\cite{Witten2002,Das2007,HohmZwiebach2019,WangWuYangYing2019}).
Moreover, it is conceptually interesting
to compare our classical degenerate-metric bounce
with the quantum bounce obtained from the
de-Broglie-Bohm pilot-wave approach
(see Ref.~\cite{Pinto-NetoFabris2013}
for a review and, in particular, the discussion of
Sec.~4.2.1 in that reference).

In closing, we remark that we have focused on the dynamics of
a time-symmetric nonsingular bounce, with equal
equation-of-state parameter $w_{M}$ in the prebounce phase and the
postbounce phase. But the metric \eqref{eq:NEW02-mod-FLRW-ds2}
is perfectly suited for the case of a
nonsingular bounce with different values of $w_{M}$ before and after
the bounce. Such a time-asymmetric
nonsingular bounce may be preferable for cosmological applications;
see Ref.~\cite{IjjasSteinhardt2018} and references therein.
The origin of the time asymmetry may be due to a fundamental
arrow of time~\cite{KlinkhamerWang2019-LHEP}  
but may also be due to dissipative processes~\cite{IjjasSteinhardt2018},
or a combination of both.

\begin{acknowledgments}
It is a pleasure to thank
J.M. Queiruga and Z.L. Wang for comments on the manuscript
and the referee for comments on an earlier version of this article.  
\end{acknowledgments}

\begin{appendix}
\section{Solutions from initial conditions}
\label{app:Solutions-from-initial-conditions}

\subsection{Solution with bounce}  
\label{subapp:Solution-with-bounce}

\begin{figure}[t]  
\vspace*{-2mm}
\begin{center}
\includegraphics[width=1\textwidth]{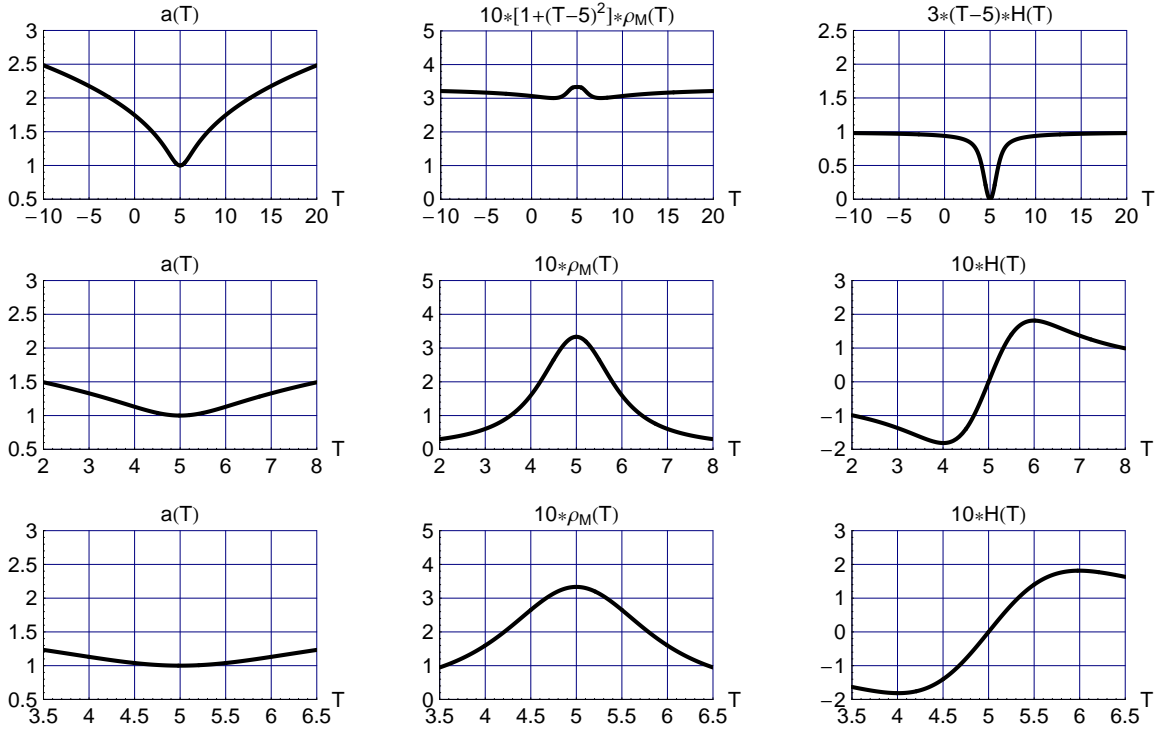}
\end{center}
\vspace*{-8mm}
\caption{Solution of the ODEs
\eqref{eq:NEW02-mod-Feq-rhoMprimeeq} and \eqref{eq:NEW02-mod-Feq-1stFeq}
for $a_{B} = 1$ and constant equation-of-state parameter $w_{M}$
from \eqref{eq:constant-wM}.
The model parameters are $b=1$ and $w_{M}=1$.
Different from Fig.~\ref{fig:fig1},
there are now initial boundary conditions at $T=T_\text{start}=-10$.
Specifically, the boundary conditions are
$a(-10)=2.48312$ and $\rho_{M}(-10)$
from \eqref{eq:initial-condition-rhoM-from-r0-astart-appA1}
for $r_{0}=1/3$ and $w_{M}=1$.
The numerical solution gives $T_{B} \approx 5.000$.
In fact, the numerical prebounce solution is obtained over
$T \in [-10,\,5-\Delta T]$,
the analytic solution over $T \in (5-\Delta T,\,5+\Delta T)$,
and the numerical postbounce solution over $T \in [5+\Delta T,\,20]$,
with $\Delta T=1/2$.
The analytic solution for $a(T)$
is given by \eqref{eq:a-power-series} with $(T)^{2n}$
on the right-hand side replaced by $(T-5)^{2n}$
and coefficients \eqref{eq:a2-a4-a6-a8-num} for $N=4$,
while the analytic solution for $\rho_{M}(T)$
follows from \eqref{eq:rhoMsol-r0-positive}.
The numerical postbounce solution has boundary conditions
$a(5+\Delta T)=a(5-\Delta T)$, $\rho_{M}(5+\Delta T)=\rho_{M}(5-\Delta T)$,
and $a^{\prime}(5+\Delta T)>0$.
Shown, on the top row, are the dynamic functions $a(T)$ and
$\rho_{M}(T)$, together with the corresponding
Hubble parameter $H(T) \equiv [d a(T)/d T]/a(T)$.
The middle and bottom rows zoom in on the bounce at $T=T_{B}\approx 5.000$.
In the middle panel of the top row, $10\,\rho_{M}(T)$ is scaled by a
further factor $\big[1+(T-5)^{2}\big]$,
in order to display the asymptotic behavior
$\rho_{M}(T) \propto  1/(T-5)^{2}$  as $|T-5|\to\infty$.
Similarly, in the right panel of the top row, $H(T)$ is scaled by a
factor $3\,(T-5)$, in order to display the asymptotic
behavior $H(T) \sim (1/3)\,(T-5)^{-1}$.
}
\label{fig:fig5}
\end{figure}

The nonsingular bounce solution
in Sec.~\ref{sec:Bounce-solution-with-bcs-at-the-bounce} was obtained
from boundary conditions at the moment of the bounce, $T=T_{B}$.
Specifically, the boundary conditions of
Figs.~\ref{fig:fig1}--\ref{fig:fig4} were
given at $T=T_{B}=0$:
$a(0) = a_{B}=1$ and $\rho_{M}(0)=r_{0}$
for values $r_{0}=1/3$ or $r_{0}=1/6$.
Here, we present a nonsingular bounce solution obtained from
\emph{initial} boundary conditions at $T=T_\text{start}$,
where the actual value of $T_{B}$ follows from the solution itself.

Consider the ODEs
\eqref{eq:NEW02-mod-Feq-rhoMprimeeq} and \eqref{eq:NEW02-mod-Feq-1stFeq}
for a constant equation-of-state parameter $w_{M}$
from \eqref{eq:constant-wM} and set
\beq
\label{eq:aB-unity-appA1}
a_{B}=1\,.
\eeq
Next, choose an arbitrary time
\beq
\label{eq:Tstart}
T_\text{start} \in \mathbb{R}\,,
\eeq
at which the initial conditions are the following:
\bsubeqs\label{eq:initial-conditions-appA1}
\beqa
a(T_\text{start}) &=& a_\text{start}  > 1\,,
\\[2mm]
\rho_{M}(T_\text{start})&=& \rho_{M-\text{start}} > 0\,,
\\[2mm]
a^{\prime}(T_\text{start}) &<& 0\,,
\eeqa
\esubeqs
where the prime stands for differentiation with respect to $T$.
If, for a generic value of $a_\text{start}$ (with $a_\text{start}>1$),
the value of $\rho_{M-\text{start}}$ is chosen as
\beq
\label{eq:initial-condition-rhoM-from-r0-astart-appA1}
\rho_{M-\text{start}}=r_{0}\;
\big(a_{B}\big/a_\text{start}\big)^{3\,\left[1+w_{M}\right]}\,,
\eeq
then, for $r_{0}=1/3$ and $w_{M}=1$, the bounce dynamics of
Figs.~\ref{fig:fig1} and \ref{fig:fig3}
is reproduced, but with $T_{B}=0$ shifted to a nonzero
value $T_{B}> T_\text{start}$.
Similarly, for $r_{0}=1/6$ and $w_{M}=1$, the bounce dynamics of
Figs.~\ref{fig:fig2} and \ref{fig:fig4} is reproduced.

The solution from
initial conditions \eqref{eq:initial-conditions-appA1}
and \eqref{eq:initial-condition-rhoM-from-r0-astart-appA1}
at $T_\text{start}=-10$ is shown in Fig.~\ref{fig:fig5}.
The actual value $a_\text{start}=2.48312$, for given values
$r_{0}=1/3$ and $w_{M}=1$
in \eqref{eq:initial-condition-rhoM-from-r0-astart-appA1},
is chosen so that the resulting value for $T_{B}$
is close to $T_\text{start}+15 \approx 5$,
which makes the curves of Fig.~\ref{fig:fig5} resemble those of
Fig.~\ref{fig:fig1}
(with more digits in the numerical value of $a_\text{start}$,
the curves of Fig.~\ref{fig:fig5} become essentially identical to
those of Fig.~\ref{fig:fig1} with a shifted $T$ coordinate).
Different values of $a_\text{start}$
(for the same values of $T_\text{start}$, $r_{0}$, and $w_{M}$)
give different values for $T_{B}$.

\subsection{Solution without bounce}  
\label{subapp:Solution-without-bounce}

We discuss, for the case of a positive cosmological constant,  
a particular analytic solution of the ODEs \eqref{eq:NEW02-mod-Feqs}
from the type-2 metric, and compare it to
the corresponding analytic solution of the ODEs \eqref{eq:previous-bounce-mod-Feqs}
from the type-1 metric. 
Even though both analytic solutions use the same type of initial conditions, 
their behavior is very different: 
the first solution does not have a bounce, whereas the second one has.

Having a positive cosmological constant $\Lambda$ corresponds to setting
\beq
\label{eq:CC-appA2}
\rho_{M}(T) =-P_{M}(T)=\Lambda=\text{constant}>0\,,
\eeq
which trivially solves \eqref{eq:NEW02-mod-Feq-rhoMprimeeq}.
The remaining ODE is given by \eqref{eq:NEW02-mod-Feq-1stFeq}
which reads for $a_{B}=1$:
\beq
\label{eq:NEW02-mod-1stFeq-CC-appA2}
\left[1+ \frac{b^{2}}{4}\,\frac{a^{2}}{\big[a-1\big]^{2}}\,
\left( \frac{a^{\prime}}{a}\right)^{2}\,\right]\,
\left( \frac{a^{\prime}}{a}\right)^{2}
= \frac{8\pi G_{N}}{3}\,\Lambda\,.
\eeq
Now take the following initial conditions:
\bsubeqs\label{eq:initial-conditions-appA2}
\beqa
\label{eq:initial-conditions-astart-appA2}
a(T_\text{start})&<&0\,,
\\[2mm]
\label{eq:initial-conditions-Tstart-appA2}
T_\text{start}   &<&0\,.
\eeqa
\esubeqs

Neglecting $b$, one solution of \eqref{eq:NEW02-mod-1stFeq-CC-appA2}
with initial conditions \eqref{eq:initial-conditions-appA2}
is simply given by
\bsubeqs\label{eq:a-sol-CC-for-zero-b-appA2}
\beqa
a(T)\,\Big|^{(b=0)}&=& -\exp \left[\, H_\text{dS}\,T\, \right]\,,
\\[2mm]
H_\text{dS} &\equiv& \sqrt{8\pi G_{N}\,\Lambda/3}\,.
\eeqa
\esubeqs
For small but nonzero $b$
[specifically, $0< (b\,H_\text{dS})^2\ll 1$],
the solution can be written as
\bsubeqs\label{eq:a-sol-CC-for-small-b-AND-f-asymptotics-appA2}
\beq\label{eq:a-sol-CC-for-small-b-appA2}
a(T)= -\exp \left[\, H_\text{dS}\,f(T)\,T \,\right]\,,
\eeq
with a smooth function $f(T)$ that remains close to $1$ and
has the following asymptotic behavior:
\beq
\label{eq:f-asymptotics-appA2}
f(T)\sim \left\{
        \begin{array}{ll}
          1, & \;\;\text{for}\;\;T\ll -b \,, \\
          (b\,H_\text{dS})^{-1}\,
         \left[ 2\,\sqrt{1+(b\,H_\text{dS})^2}-2 \right]^{1/2}\,,
            & \;\;\text{for}\;\;T\gg b \,.
        \end{array}
      \right.
\eeq
\esubeqs
The solution $a(T)$
from \eqref{eq:a-sol-CC-for-small-b-AND-f-asymptotics-appA2}
is monotonic and does not display bounce behavior at any finite value of $T$.
Incidentally, a bounce solution does occur if, for example,
the initial condition \eqref{eq:initial-conditions-astart-appA2}
is replaced by $a(T_\text{start})>1$:
with negative $a'/a$ at $T=T_\text{start}$,
the solution has a bounce at $T_{B}>T_\text{start}$
and qualitatively resembles the solution found in
Appendix~\ref{subapp:Solution-with-bounce},
whereas, with positive $a'/a$ at $T=T_\text{start}$,
the solution has a bounce at $T_{B}<T_\text{start}$.

Returning to the solution \eqref{eq:a-sol-CC-for-small-b-AND-f-asymptotics-appA2}
with initial conditions \eqref{eq:initial-conditions-appA2},
the corresponding solution for the type-1 metric \eqref{eq:mod-FLRW}
has been discussed in Appendix~B of Ref.~\cite{Klinkhamer2019-PRD}.
With similar initial conditions,
$\widetilde{a}(T_\text{start})<0$ for $T_\text{start}   <0$,
such a solution of the ODE \eqref{eq:previous-bounce-mod-Feq-1stFeq}
with $\rho_{M}=\Lambda>0$ is given by%
\beq
\label{eq:mod-FLRW-CC-bounce-sol-appA2}
\widetilde{a}(T) = -\exp \left[ - H_\text{dS}\,\sqrt{b^2+T^2}\, \right]\,,
\eeq
which displays a nonsingular bounce behavior at $T=0$  for $b\ne 0$.
Incidentally, a similar bounce solution occurs
if the initial condition $\widetilde{a}(T_\text{start})>0$
is used: the corresponding solution is then given
by \eqref{eq:mod-FLRW-CC-bounce-sol-appA2}
with a plus sign in front of the exponential function on the right-hand side.
Other bounce solutions, with
\beq
\label{eq:mod-FLRW-CC-bounce-other-sol-appA2}
\widetilde{a}(T) = \mp \exp \left[ H_\text{dS}\,\sqrt{b^2+T^2}
-2\, H_\text{dS}\,\sqrt{b^2+(T_\text{start})^2}\, \right]\,,
\eeq
may be more relevant for cosmology,
as $\widetilde{a}^2(T) \to \infty$ for $|T|\to \infty$.

The different solutions \eqref{eq:a-sol-CC-for-small-b-AND-f-asymptotics-appA2}
and \eqref{eq:mod-FLRW-CC-bounce-sol-appA2}
make clear that the corresponding metrics \eqref{eq:NEW02-mod-FLRW}
and \eqref{eq:mod-FLRW} are essentially different.
For both spacetime metrics,
the Cauchy problem~\cite{HawkingEllis1973}
and the foliation dependence or independence  
\cite{Teitelboim1973,HojmanKucharTeitelboim1976} deserve further study.

\section{Comparison with loop quantum cosmology and string cosmology}
\label{app:Comparison-with-loop-quantum-cosmology-and-string-cosmology}

In this appendix, we review the bounce of the  cosmic scale factor
obtained from loop quantum cosmology (LQC) and compare with the
bounce obtained from extended general relativity with
the degenerate metric \eqref{eq:NEW02-mod-FLRW}. For the record,
we also compare with the bounce of string cosmology (SC).  
We set $c=1$ and $\hbar=1$, but keep $G_{N}$ explicit.

As shown in Appendix~B 1 of Ref~\cite{AshtekarSingh2006}
(and discussed in later reviews~\cite{AshtekarSingh2011,AshtekarBarrau2015}), 
the LQC bounce can be described by an
\emph{effective} Friedmann equation,
\bsubeqs\label{eq:LCC-eff-Feq-rhoB}
\beqa
\label{eq:LCC-eff-Feq}
\left( \frac{\dot{a}}{a}\right)^{2}
&=&
\frac{8\pi G_{N}}{3}\,\rho\,\left( 1 - \frac{\rho}{\rho_{B}} \right)\,,
\\[2mm]
\label{eq:LCC-rhoB}
\rho_{B} &=& c_{B}\,\big(E_\text{Planck}\big)^{4}\,,
\eeqa
\beqa
\label{eq:LCC-E-Planck}
E_\text{Planck} &\equiv&  1/\sqrt{G_{N}}
\approx 1.22 \times 10^{19}\,\text{GeV}\,,
\\[2mm]
\label{eq:LCC-l-Planck}
l_\text{Planck} &\equiv& 1/E_\text{Planck}
\approx 1.62 \times 10^{-35}\,\text{m}\,,
\eeqa
\esubeqs
where the overdot stands for the derivative with respect to
the cosmic time coordinate $t \in \mathbb{R}$.
The dimensionless constant $c_{B}$ in \eqref{eq:LCC-rhoB}
is positive and, most likely, of
order unity. In physical terms~\cite{AshtekarSingh2011,AshtekarBarrau2015},
the value of $\rho_{B}$ is set by the area gap
$\Delta \sim (l_\text{Planck})^{2}$, $\rho_{B} \sim 1/\Delta^{2}$.
Let us have a quick look at two domains of this
bouncing cosmology, far away from the bounce and close to it.

For energy densities $\rho\ll \rho_{B}$
(or cosmic scale factors $a \gg a_{B} \equiv 1$),
the effective Friedmann equation \eqref{eq:LCC-eff-Feq-rhoB}
can be rewritten as
\beq
\label{eq:LCC-eff-Feq-far-from-bounce}
\left[1+ \frac{3}{8\pi G_{N}}\,\frac{1}{\rho_{B}}\,
\left( \frac{\dot{a}}{a}\right)^{2}\,\right]\,
\left( \frac{\dot{a}}{a}\right)^{2}
\sim \frac{8\pi G_{N}}{3}\,\rho\,.
\eeq
The structure of \eqref{eq:LCC-eff-Feq-far-from-bounce}
is the same as the one of our modified
Friedmann equation \eqref{eq:NEW02-mod-Feq-1stFeq}
with $a^2/[a - a_{B}]^2 \sim 1$.
This allows for the following tentative identification  
\beq
\label{eq:b-square-from-LQC}
\frac{b^{2}}{4} \;\;\stackrel{\text{LQC?}}{\sim}\;\;
\frac{3}{8\pi G_{N}}\,\frac{1}{\rho_{B}}
=
\frac{3}{8\pi}\,\frac{(E_\text{Planck})^{2}}{\rho_{B}}\,,
\eeq
so that $b$ would be of the order of $l_\text{Planck}$,
as long as $c_{B}$ from \eqref{eq:LCC-rhoB} is of order unity.

For energy densities $\rho$ close to $\rho_{B}$,
it is possible to use a series \textit{Ansatz},
\bsubeqs\label{eq:LCC-a-series-rho-series}
\beqa
a(t) &=& 1+ \widehat{a}_{2}\,t^{2} + \ldots \,,
\\[2mm]
\rho(t) &=& \rho_{B}+ \widehat{r}_{2}\,t^{2} + \ldots \,,
\eeqa
\esubeqs
with $\rho_{B}>0$ and $\widehat{r}_{2} <0$.
Inserting  \eqref{eq:LCC-a-series-rho-series} into
\eqref{eq:LCC-eff-Feq} gives
\beq
\label{eq:LCC-widehat-a-2-result}
(\widehat{a}_{2})^{2} = - \frac{2\pi G_{N}}{3}\,\widehat{r}_{2}\,.
\eeq
The quadratic coefficient $\widehat{a}_{2}$ from
\eqref{eq:LCC-widehat-a-2-result} has no direct dependence on $\rho_{B}$,
which is given by fundamental constants
according to \eqref{eq:LCC-rhoB}.
This result differs from what has been obtained
from the degenerate-metric bounce, where the quadratic
coefficient \eqref{eq:a2} depends on $\sqrt{r_0}$, with
$r_0 = 8\pi G_{N}\,\rho_{B}$ being a free parameter.
A further difference is that, 
for the case of relativistic matter ($P=\rho/3$),  
the Ricci curvature scalar at the moment
of the bounce does not vanish in the LQC calculation,
according to Eq.~(5.13) of Ref.~\cite{AshtekarSingh2011},
whereas it does vanish in the extended-general-relativity calculation,
according to \eqref{eq:R0} of Sec.~\ref{subsec:Analytic-results}.

The conclusion is that the degenerate metric
\eqref{eq:NEW02-mod-FLRW}  may give a reasonable
approximation of the LQC-bounce behavior at relatively low energy densities
but not of the LQC-bounce dynamics close to the maximum energy density.
At this moment, it is not clear which of the two models,
extended general relativity with an appropriate degenerate metric
or the effective theory from loop quantum cosmology, best describes
the cosmic bounce, assuming such a bounce to be relevant to our
Universe.

But another possibility is that the degenerate-metric bounce
results from an entirely new phase~\cite{Witten2002} of string theory;
see, e.g., Ref.~\cite{Das2007} for a further review.
For a direct comparison, we refer to two recent 
string-cosmology  
calculations~\cite{HohmZwiebach2019,WangWuYangYing2019}.

The authors of Ref.~\cite{WangWuYangYing2019}
obtained an explicit nonperturbative solution of the
reduced equations of motion, assuming
certain higher-order coefficients $c_k$, for $k \geq 3$,
in the effective one-dimensional action of Ref.~\cite{HohmZwiebach2019}.
Specifically, the Hubble parameter $H_{+}^{E}(t)$
in the Einstein frame, for spatial dimensionality $d=3$,
is given by Eq.~(3.24) in Ref.~\cite{WangWuYangYing2019}
and has the following structure:
\bsubeqs\label{eq:H-t0-string-cosmology}
\beqa
\label{eq:H-string-cosmology}
H_{+}^{E}(t) &\sim&
\frac{t-t_{B,+}}{\alpha^{\prime}/6+t^2}\,,
\\[2mm]
\label{eq:t0-string-cosmology}
t_{B,+} &\sim& - \sqrt{\alpha^{\prime}}\,,
\eeqa
\esubeqs
where $\alpha^{\prime}>0$ is the Regge slope
related to the inverse of the string tension
and $t_{B,+}$ the moment of the bounce
\big[the other solution $H_{-}^{E}(t)$ is also
given by \eqref{eq:H-string-cosmology}, but with $t_{B,+}$
replaced by $t_{B,-} \sim \sqrt{\alpha^{\prime}}\;$\big].
The Hubble parameter $H_{+}^{E}(t)$
is shown in Fig.~4 of Ref.~\cite{WangWuYangYing2019}
and resembles qualitatively the Hubble parameter
obtained from the degenerate metric
\eqref{eq:NEW02-mod-FLRW}, 
as shown by the mid-right panels  
in \mbox{Figs.~\ref{fig:fig1} and \ref{fig:fig2}.}

Comparing the generic degenerate-metric result $H(t) \sim t/(b^2+t^2)$
from Ref.~\cite{Klinkhamer2019-PRD} with \eqref{eq:H-string-cosmology}
we have the following tentative identification:
\beq
\label{eq:b-square-from-string-cosmology}
b^{2} \;\;\stackrel{\text{SC?}}{\sim}\;\;\frac{1}{6}\; \alpha^{\prime}\,,
\eeq
which may be of order $(l_\text{Planck})^2$.
Still, it needs to be emphasized that the existing
string-cosmology 
calculations (just as the existing loop-quantum-cosmology calculations)
are only indicative and that a possible string-theory phase replacing the
big bang singularity may have highly unusual
properties~\cite{Witten2002}.

\end{appendix}


\end{document}